\documentclass[%
 reprint,
 amsmath,amssymb,
 aps,
floatfix,
]{revtex4-1}

\usepackage{graphicx}
\usepackage{dcolumn}
\usepackage{bm}
\usepackage{subfigure}
\usepackage{hyperref}
\usepackage{wrapfig}
\usepackage{placeins}
\usepackage{natbib}

\begin{document}


\title{The Origin of Tilted Phase Generation in Systems of Ellipsoidal Molecules with  Dipolar Interactions}


\author{Tushar Kanti Bose}
\email{tkb.tkbose@gmail.com}
\author{Jayashree Saha}
\email{jsphy@caluniv.ac.in, corresponding author}
\affiliation{Department of Physics, University of Calcutta, 92 A.P.C. Road, Kolkata-700 009.}
 




\date{\today}

\begin{abstract}
    We report Monte-Carlo simulation studies of some systems consisting of
polar rod-like molecules interacting via a pair potential that
exhibit liquid crystal phases, attributed with tilt angles of large magnitude. For theoretical understanding of the microscopic origin of
the tilted phases, different systems consisting of prolate ellipsoidal molecules of three different
lengths, embedded with two symmetrically placed anti-parallel terminal dipoles
are considered. We find that the presence of a stable tilted phase crucially
depends on the molecular elongation which effectively makes dipolar separation
longer. We observe that in case of mesogens with transverse dipoles the tilt
in the layered smectic phase gradually increases from zero to a large magnitude as we increase
the molecular length. However tilt remains weak with molecular elongation for
systems with longitudinal dipoles which shows a small tilt at shorter lengths. This is the first work determining the combined contribution of dipolar separation and orientations in generating biaxial liquid crystal phases with large tilt angles.





\end{abstract}

\pacs{Valid PACS appear here}
\maketitle



Polar liquid crystals exhibit a rich variety of tilted layered mesophases
arising out from the diversity in both the inlayer positional arrangements of the molecules
and in the average molecular orientations with respect to the layer normal \cite{jon1,jon2}.
The most common example of such a tilted Smectic phase is the Smectic C ( SmC ) phase where the center of mass of the tilted molecules are randomly 
distributed in two dimensional fluid like layers alongwith a long range orientational ordering of molecules over the bulk system. Tilted phases with the same point symmetry as SmC
phase also exist in other orientationally ordered soft matter
systems like mesogenic polymers, lamellar L\(_{\beta}\) phase and in SmC elastomers \cite{jon3,jon4,jon5}.
The tilted phases are important
not only from the fundamental scientific viewpoint but recent discoveries of 
a large variety of ferroelectric, anti-ferroelectric and ferrielectric phase behavior in the tilted phases have
made them a topic of huge interest from technological point of view. These phases are used for developing new generation opto-electric devices and various
non-display applications \cite{jon6}. 


The origin of 
tilt in liquid crystalline phases has been a topic of much discussion.
A large number of theoretical and experimental studies have been
done on tilted phases. However the topic remains unsolved in certain aspects till today.
It is essential to understand the structure-property relationship 
to find the basic interactions
giving rise to tilt in Sm C phase. The microscopic origin of the
 tilted smectic phases are much more complex than that of the orthogonal smectic
 phases since the existence of a tilted phase is not a favored one because 
packing of tilted rod like molecules in a layer plane needs more area
than the untilted molecules. This phase can be achieved only if there
exists additional specific interactions giving the requisite tilt. 

In the experimental studies, occurrence of the SmC phases
has been mostly found in presence of a lateral component of permanent dipole 
moment in the organic molecules \cite{jon1,jon2}. As the number of such dipoles
increases on a molecule, the tendency to form SmC phase also increases \cite{jon7}.
Motivated by the experimental 
observations, Mcmillan gave the mean field theory on the formation of SmC phase based 
on the presence of at least two outboard terminal dipoles \cite{jon8}. However a freezing of
rotation of the molecules was resulted during such a SmA to SmC phase transition contradicting
 NMR \cite{jon9} and neutron scattering experiments \cite{jon10} which have shown the free rotation of molecules
in a tilted smectic phase. Wulf attempted to relate the formation of SmC
phase to the packing requirements of the zig-zag shaped molecules \cite{jon11}. However lowering
of free energy was associated to a freezing of rotations. A zig-zag model made of
seven Lennard Jones spheres with two terminal ones at an angle \(45^{\circ}\) from the
five in line core, showed a SmC behavior but the tilt orientation was random and
the equilibration process was elaborate \cite{jon12}. A model based on three rigidly linked hard
spherocylinders arranged in a zig-zag fashion have shown the presence of SmC phase \cite{jon13}.
However the simple model particles built by assembling in a zig-zag way ellipsoidal Gay-Berne
particles was unable to show tilted phases \cite{jon14}. In computer simulation studies by Zannoni et.al. \cite{jon15},
 the axial component rather than the transverse component of lateral dipole 
moments were found to be generating a tilted phase for polar GB molecules. In their system, the biaxial order parameters were very small
and weak non zero tilt was found for the system of molecules with lateral dipole moments making an angle \(\phi= 0^{\circ}\) or \( 60^{\circ}\) with the long axis. Similar results were found by Saha et.al.\cite{jon16} for a large dipolar oriention \(\phi=120^{\circ}\). However absence of tilt in these conventional model of polar molecules with transverse dipoles, contradicting experimental results, remains an open problem.

We explored systems of polar ellipsoidal molecules where each molecule is embedded with two terminal
 anti parallel permanent dipole moments placed at equal distance from the center of the molecules to get considerable
 tilt in the smectic phases. In our model, the apolar part of the interaction is represented by the Gay-Berne(GB) potential which is a modified form 
of the Lennard-Jones potential, considering the anisotropy in interaction.
The interaction between two such apolar GB ellipsoids
i and j is given by
\[ U_{ij}^{GB}(\mathbf{r}_{ij},\hat{u}_{i},\hat{u}_{j})=4\epsilon(\hat{r}_{ij},\hat{u}_{i},\hat{u}_{j})(\rho_{ij}^{-12}-\rho_{ij}^{-6}) \]

\(\displaystyle{\mbox{   where  }\displaystyle{\rho_{ij} = \frac{[ r_{ij} - \sigma( \hat{ r}_{ij},\hat{u}_{i},\hat{u}_{j} ) + \sigma_{0}  ] }{ \sigma_{0}} }\mbox{.}}\)
Here \(\mathbf{r}_{ij}\) is the separation vector between the center of mass of the molecules. The unit vectors \(\hat{u}_{i}\mbox{ and }\hat{u}_{j}\) represent the orientations of the molecules. \(\sigma_{0}\) is the minimum separation for a side-by-side pair of molecules determining the breadth of the molecules.  The minimum separation for an end-to-end pair of molecules \(\sigma_{e}\) is a measure of the length of the molecules. The anisotropic contact distance \(\sigma(\hat{r}_{ij},\hat{u}_{i},\hat{u}_{j})\) and the depth of the
interaction well \(\epsilon(\hat{r}_{ij},\hat{u}_{i},\hat{u}_{j})\) depend on the the shape anisotropy parameter \(\kappa=\sigma_{e}/\sigma_{0}\) and the energy depth anisotropy parameter    \(   \kappa^{'} = \epsilon_{e} / \epsilon_{s}  \) which are defined as the ratios of the contact distances and energy well depths in the end-to-end and side-by-side configurations. The anisotropic contact distance varies with \(\kappa\) as \(\sigma(\hat{r}_{ij},\hat{u}_{i},\hat{u}_{j}) \) is given by \(\sigma=\)

\(\displaystyle{    \sigma_{0}  \left[ 1 - \frac{\chi}{2}\left(     \frac{(\hat{r}_{ij}.\hat{u}_{i}+\hat{r}_{ij}.\hat{u}_{j})^{2}}{1+\chi(\hat{u}_{i}.\hat{u}_{j})} +  \frac{(\hat{r}_{ij}.\hat{u}_{i}-\hat{r}_{ij}.\hat{u}_{j})^{2}}{1-\chi(\hat{u}_{i}.\hat{u}_{j})}                   \right)  \right]^{-\frac{1}{2}     }}\)\\

where \(\chi=(\kappa^{2}-1)/(\kappa^{2}+1)\). The anisotropy of the well depth \(\epsilon\) is also controlled by two additional parameters \(\mu\mbox{ and }\nu \). An explicit description of the GB interaction can be found in the original paper \cite{jon17}.
The well depth in the cross configuration is written as \(\epsilon_{0}\). \( \sigma_{0}\mbox{ and }\epsilon_{0}  \) define the length and energy scales. We have used reduced units in
 our calculations by expressing lengths and interaction energies in units of \( \sigma_{0}\mbox{ and }\epsilon_{0}  \) respectively.
In the present work, we put two point dipole monents on each GB molecule at a reduced distance  \( d^{*} = (\kappa-1)/2  \) along the symmetry axis from the center of mass of the uniaxial molecules with shape anisotropy \(\kappa\). We have studied the bulk phase behavior for three different values of \(\kappa\)(3, 4 and 5) keeping the other parameters fixed to their original values \(\kappa^{'}=5, \mu=2, \nu=1\) in order to investigate the effects of varying the dipolar separation to large values.
\begin{figure}
\subfigure[ ]{\includegraphics[scale=0.25]{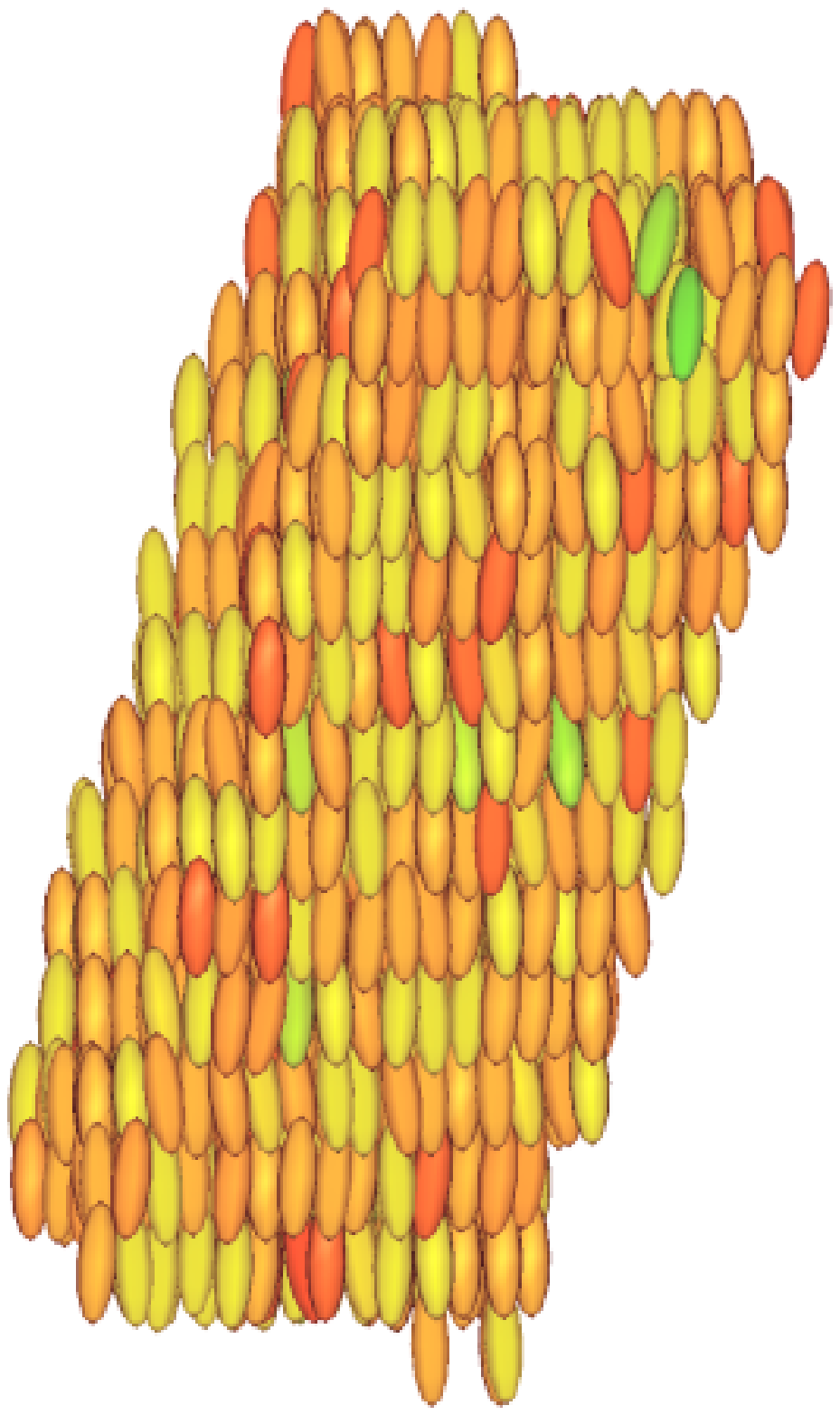}}\hspace{-0.2cm}
\subfigure[\small{}]{\label{nematic}\includegraphics[scale=0.25]{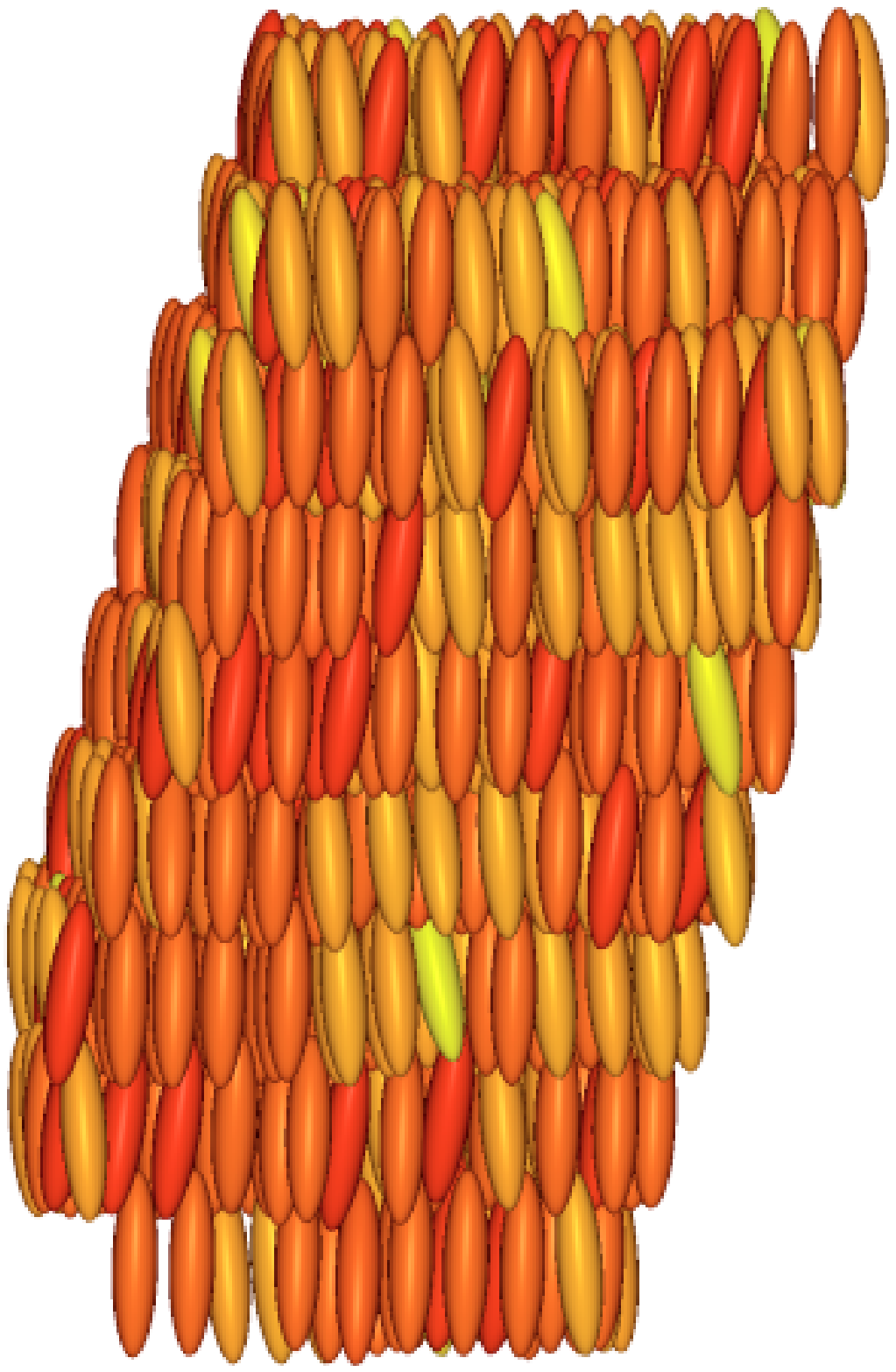}}\hspace{-0.2cm}
\subfigure[]{\includegraphics[scale=0.25]{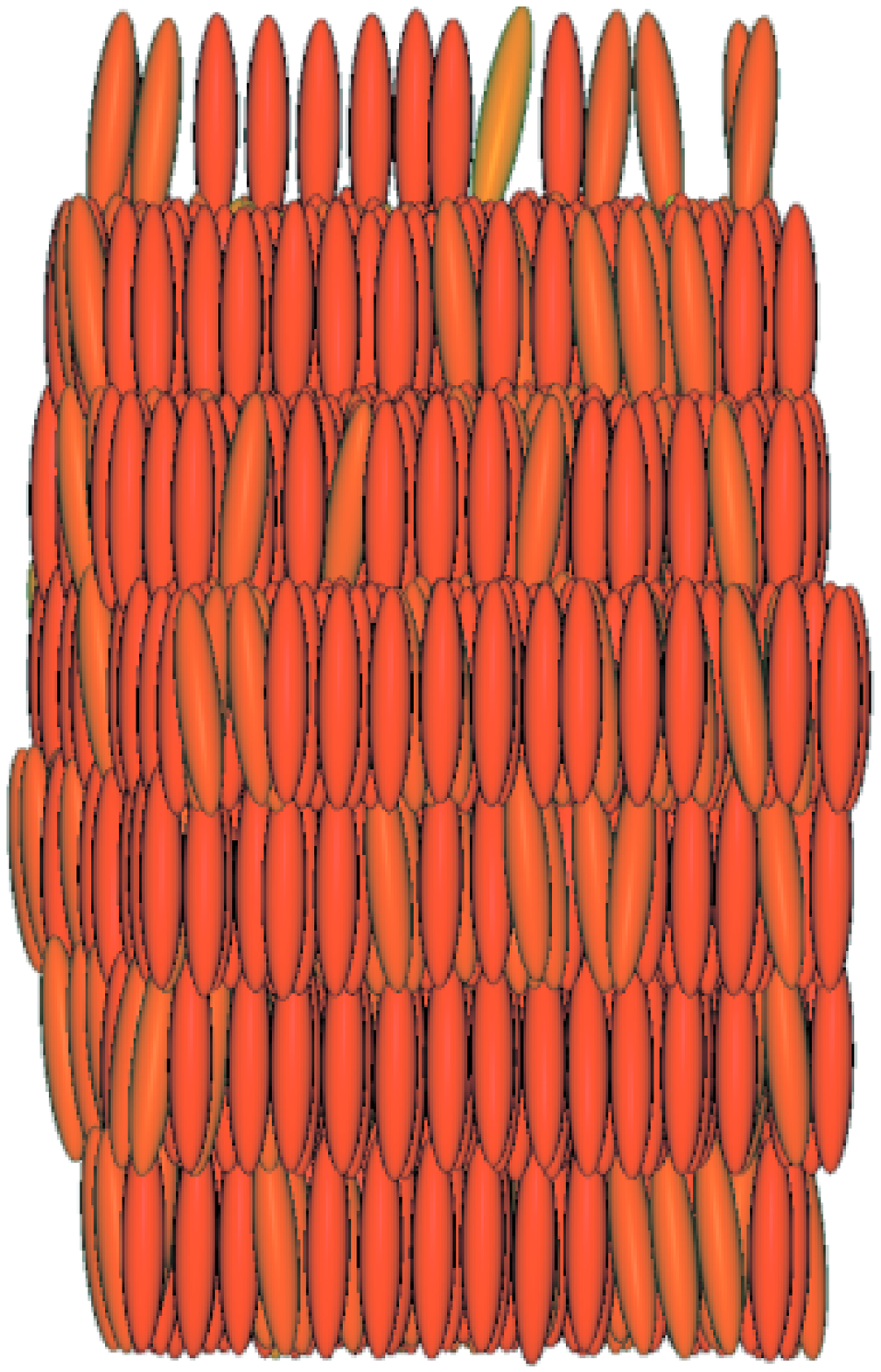}}
\caption{\label{fig:wide1}(color online). Snapshots of the final configurations from MC
simulations of a system of 1372 GB molecules with two terminal longitudinal dipoles
for various elongations \(\kappa\) : (a) Smectic at (\(\kappa=3,T^{*}=1.00,P^{*}=2.75\)) with
\(\langle\theta\rangle=1.6^{\circ}\),(b) Smectic at (\(\kappa=4,T^{*}=1.25,P^{*}=1.50\)) with \(\langle\theta
\rangle=0.6^{\circ}\),(c) Smectic at
 (\(\kappa=5,T^{*}=1.5,P^{*}=1.10\)) with \(\langle\theta\rangle =1.6^{\circ}\).
}
\end{figure}
\begin{figure*}
\hspace{-1.1cm}
\subfigure[ ]{\includegraphics[scale=0.22]{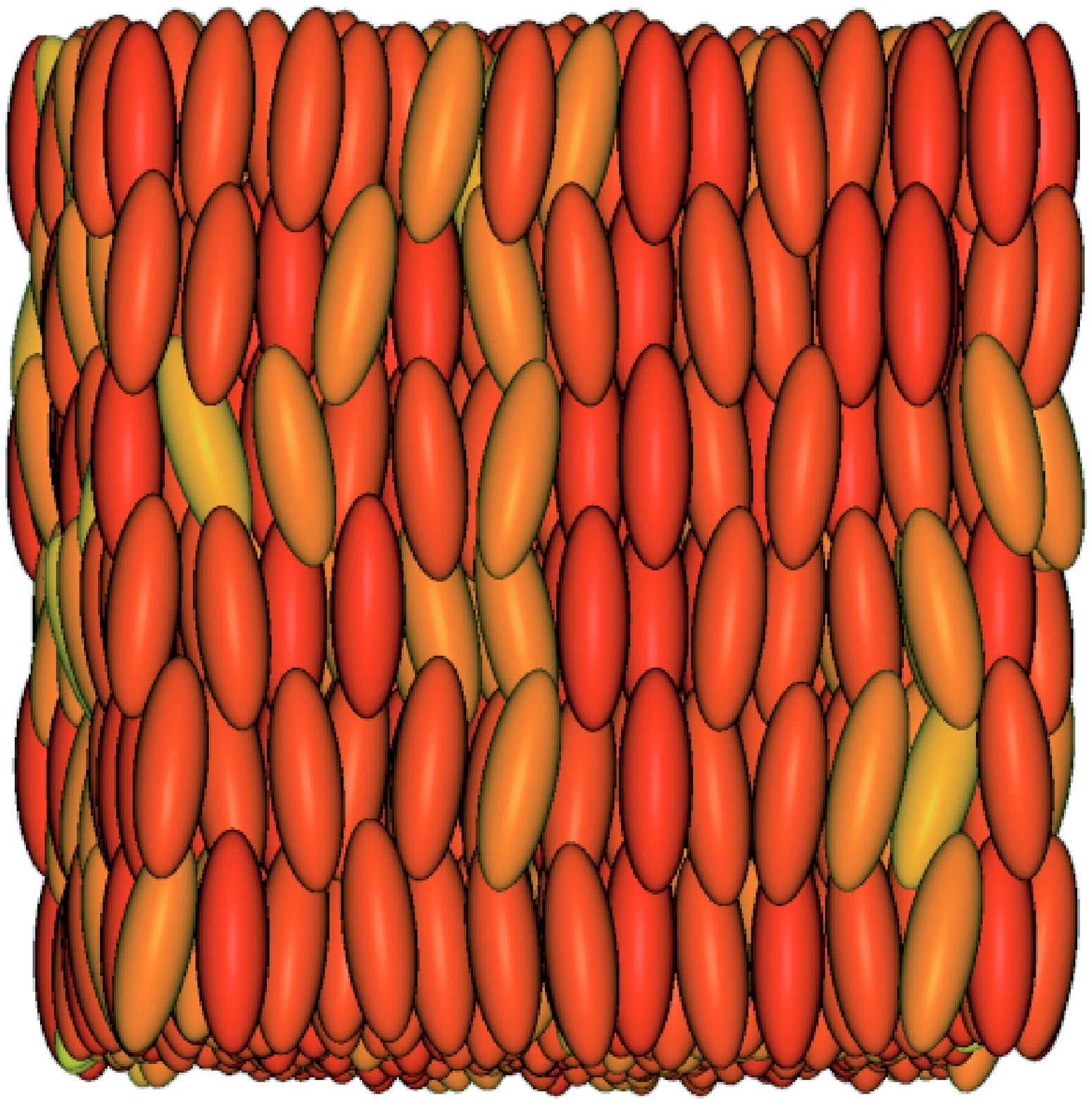}}\hspace{0.9cm}
\subfigure[\small{}]{\label{nematic}\includegraphics[scale=0.25]{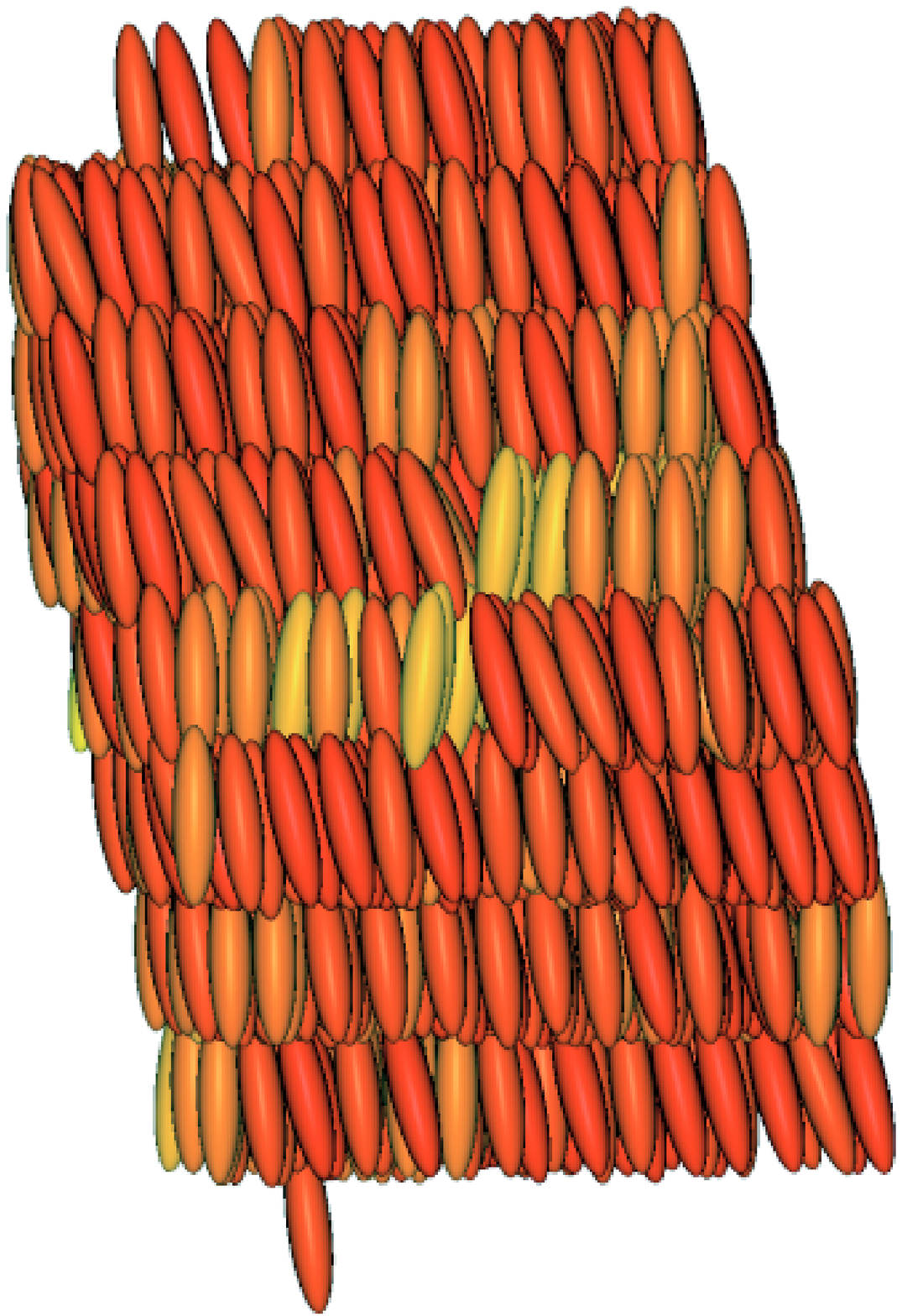}}\hspace{0.1cm}
\subfigure[]{\includegraphics[scale=0.24]{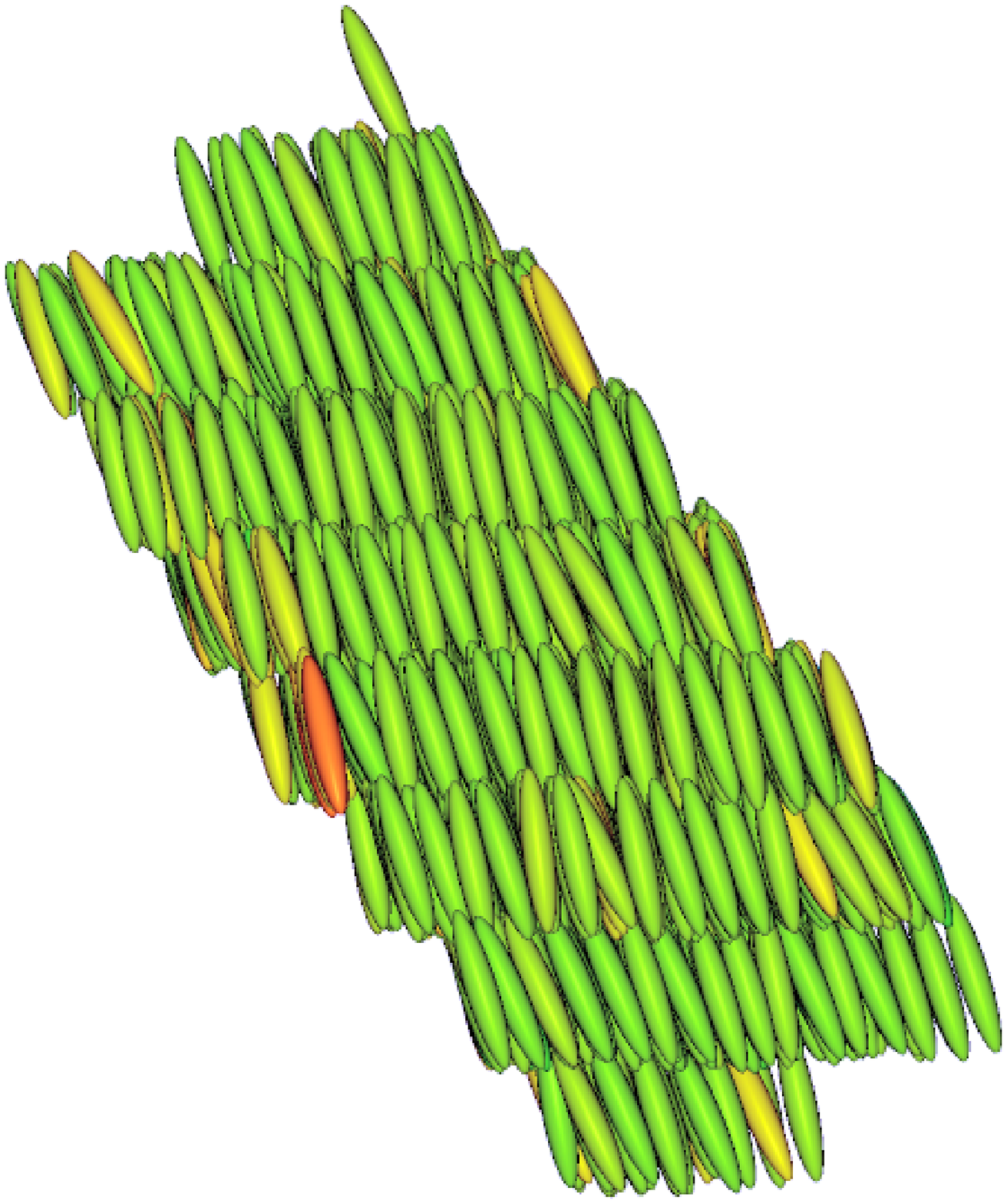}}
\subfigure[]{\includegraphics[scale=0.22]{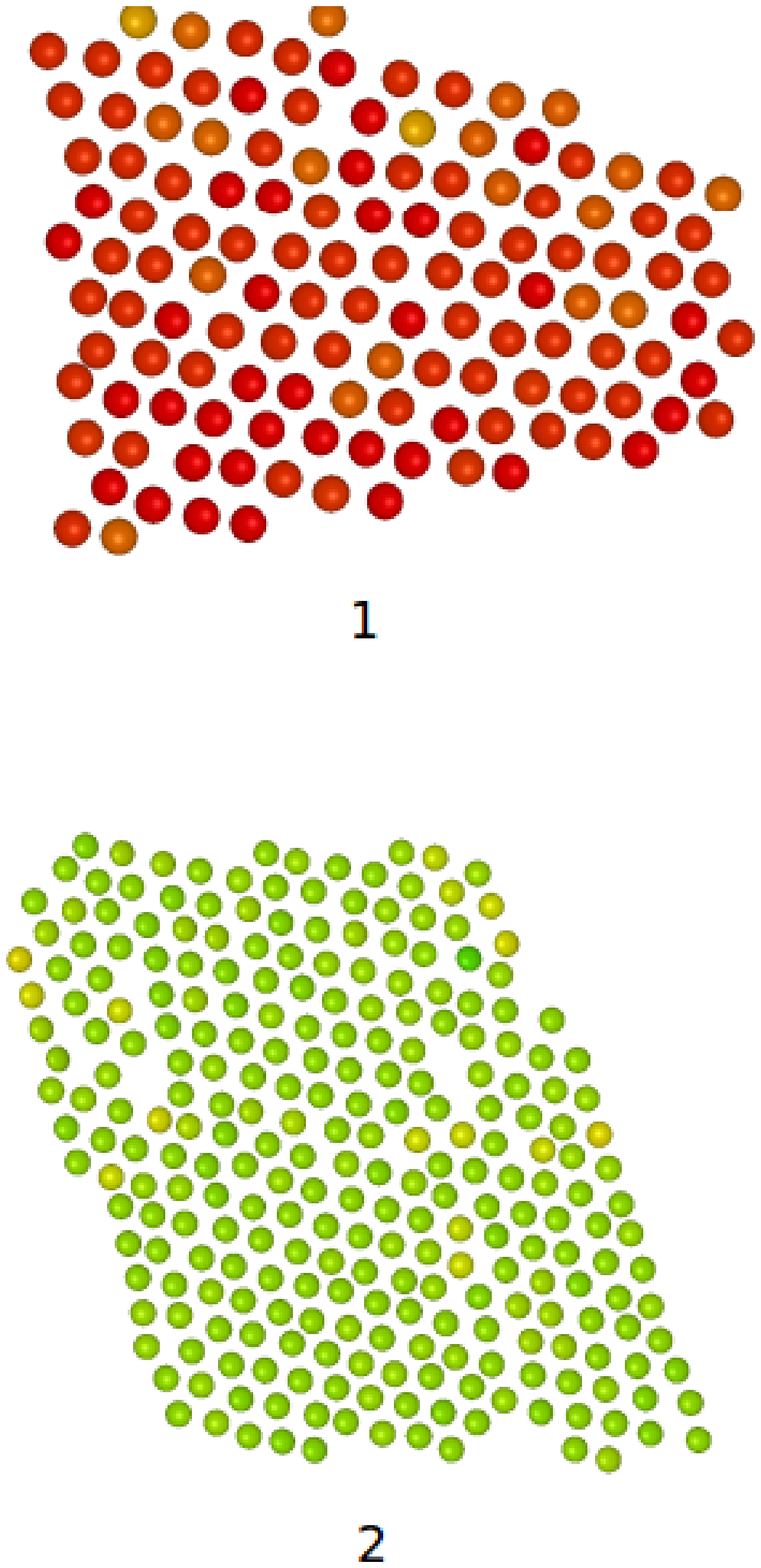}}
\caption{\label{fig:wide}(color online). Snapshots of the final configurations generated by MC
simulations of a system of 1372 GB molecules with two terminal transverse dipoles
for various elongations \(\kappa\) : (a) Smectic at (\(\kappa=3,T^{*}=1.00,P^{*}=1.35\)) with
\(\langle\theta\rangle=0.25^{\circ}\),(b) Tilted Smectic at (\(\kappa=4,T^{*}=1.25,P^{*}=1.05\)) with \(\langle\theta
\rangle=6.6^{\circ}\),(c) Tilted Smectic at
 (\(\kappa=5,T^{*}=1.5,P^{*}=1.5\)) with \(\langle\theta\rangle =17^{\circ}\),(d) molecular center of mass positions on a single 2D layer of the tilted phases showing local hexagonal orderings (1)(\(\kappa=4,T^{*}=1.25,P^{*}=1.05\)), (2)(\(\kappa=5,T^{*}=1.5,P^{*}=1.5\)).
}
\end{figure*}
The electrostatic interaction energy between two such dipolar ellipsoids is given by 
\[  {U_{ij}^{d'}}^{*}=\sum_{\alpha, \beta=1}^{2}\frac{{\mu^{*}}^{2}}{ {r_{\alpha \beta}^{*}}^{3}}\left[(\hat{\mu}_{i \alpha}.\hat{\mu}_{j \beta})-3(\hat{\mu}_{i \alpha}.\hat{r}_{\alpha \beta})(\hat{\mu}_{j \beta}.\hat{r}_{\alpha \beta})   \right]           \]where \(\textbf{r}^{*}_{\alpha \beta} (=\textbf{r}^{*}_{j \beta}-\textbf{r}^{*}_{i \alpha}) \) are the vectors joining the two point dipoles \(  \boldsymbol{\mu}^{*}_{i\alpha}   \) and \(  \boldsymbol{\mu}^{*}_{j\beta}   \) on the molecules i and j at the positions \(\textbf{r}^{*}_{i \alpha} = \textbf{r}^{*}_{i}\pm d^{*}\hat{u}_{i}\) and \(\textbf{r}^{*}_{j \beta} = \textbf{r}^{*}_{j}\pm d^{*}\hat{u}_{j}\). The reduced dipole moment \( \mu^{*} \equiv \sqrt{\mu^{2}/\epsilon_{0}\sigma_{0}^{3}}    \) is chosen \(\mu^{*}\)=1.0. The long range nature of the dipolar interaction is taken into account with the reaction field method \cite{jon18}. The dipolar part of the total interaction energy including long range correction can then be written as
\[ {U_{ij}^{d}}^{*}={U_{ij}^{d'}}^{*} -\sum_{\alpha, \beta=1}^{2}\frac{2(\epsilon_{RF}-1)}{2\epsilon_{RF}+1}\frac{\boldsymbol{\mu}^{*}_{i\alpha} \boldsymbol{\mu}^{*}_{j\beta} }{{R_{c}^{*}}^{3}} \mbox{  .}           \]Where \( R^{*}_{c}\equiv R_{c}/\sigma_{0}  \) is the reduced radius of the RF cut-off sphere and \(\epsilon_{RF}=\infty\) is the dielectric constant of the medium. Then
the total interaction between two dipolar molecules is given by \( { U_{ij}^{total}}^{*}= {U_{ij}^{GB}}
^{*}+{U_{ij}^{d}}^{*}   \).\\
We have performed Monte Carlo (MC) simulations in the NPT(isothermal-isobaric) ensemble with periodic
boundary conditions imposed on a system of N=1372 dipolar molecules. The simulation cell is an orthogonal
box of dimensions \( L_{x}\),\(L_{y}\),\(L_{z}\). The dimensions are varied
independently during simulation so that the system may fit itself to its most suitable configuration
at each state point(\(P^{*}\equiv P\sigma_{0}^{3}/\epsilon_{0},T^{*}\equiv K_{B}T/\epsilon_{0}  \)). 
All the systems are prepared initially in a completely disordered isotropic phase in a cubic box
by melting a crystal structure at sufficiently low pressure. We then increase the pressure successively by steps of \(\Delta P^{*}=\)0.10 or less (near a transition).
In each case, at a given pressure, the final equilibrated configuration obtained from previous lower pressure is used as the initial
configuration.
At each state point, the system
is equilibrated for \(3\times10^{5}\) MC cycles and \(3\times10^{6}\) MC cycles are used for equilibration near a transition.
During each MC cycle each molecule is randomly displaced and reoriented using metropolis criteria where the
reorientation moves were performed using Barker-Watts technique\cite{jon18}. One of the three box-sides was
attempted to change during each MC cycle. The acceptance rates of the roto-translational moves of
molecules and volume moves were adjusted to 40\%. 

In order to fully characterize different phases of the system various order parameters were
computed. The average orientational ordering is determined from the second-rank tensorial order parameter \(Q_{\alpha \beta}\) defined as \(Q_{\alpha \beta}=\frac{1}{N}\sum_{i=1}^{N}(\frac{3}{2}u_{i\alpha}u_{i\beta}-\frac{1}{2}\delta_{\alpha \beta})\) where \(\alpha,\beta=x,y,z\) and \(\hat{u}_{i}\) is the molecular end-to-end unit vector of molecule i. The nematic order parameter S is given by the largest eigenvalue of the ordering tensor \(Q_{\alpha \beta}\) and the corresponding eigenvector defines the phase director. The value of S is close to zero in the isotropic phase and tends to 1 in the highly ordered phase.

We have investigated the smectic structures over various elongations \( \kappa=3,4 \mbox{ and } 5 \) respectively at fixed temperatures \( T^{*}=1.00,1.25 \mbox{ and }1.50\). At such temperatures we observed a jump in S directly from \(S\sim 0.10\mbox{ to }S\sim 0.95\) indicating a direct isotropic to tilted smectic transition. To measure the average tilt angle \(\langle\theta\rangle\) of the phase director about the layer normal, we have used
the method described in \cite{jon16} i.e. in smectic phases we first find the sets of particles for which first neighbor distance is \(\leq1.3\sigma_{0}\).
Each such set forms a different layer. Then we find the normal to each such layer by a least square method. The average normal is obtained by averaging the components over all the layers of a single MC configuration. The angle \(\theta\) between the phase director and the layer normal determines the tilt. 
Then we measure its average value over a number of configurations. Since all the tilted phases are expected to be biaxial we also measure the
biaxial order parameter  \( \langle R_{2,2}^{2}\rangle=\langle\frac{1}{2}(1+\cos^{2}\beta) \cos 2\alpha \cos 2\gamma -\cos\beta \sin2 \alpha \sin2 \gamma \rangle     \) as described in \cite{jon19},where \(\alpha,\beta,\gamma \) are the Euler angles giving the orientation of the molecular body set
of axes w.r.t. the laboratory set of axes. To understand the structure of the phases we also calculate the radial distribution function \(g(r)=\frac{1}{4\pi r^{2}\rho}\langle\delta(r-r_{ij})\rangle_{ij}\mbox{ ,}\)where the average is taken over all the molecular pairs.



\begin{figure*}\hspace{-1cm}
\subfigure[\label{fig:evo_tilt1}]{\includegraphics[scale=0.9]{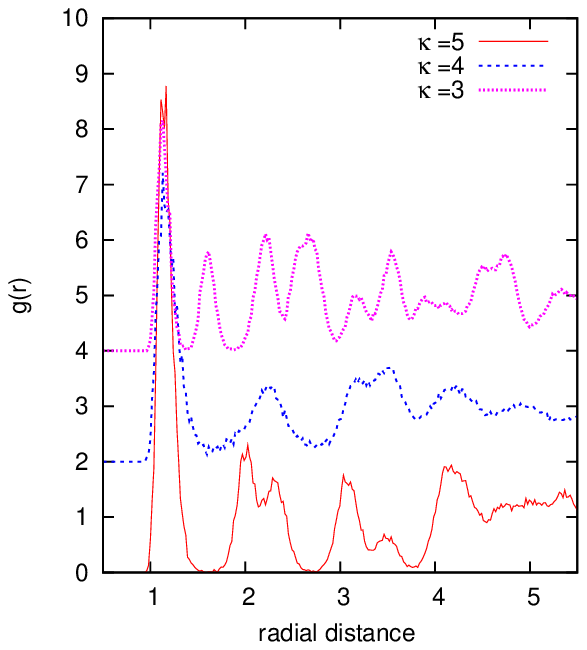}}\hspace{-0.3cm}
\subfigure[\label{fig:evo_tilt2}]{\includegraphics[scale=0.9]{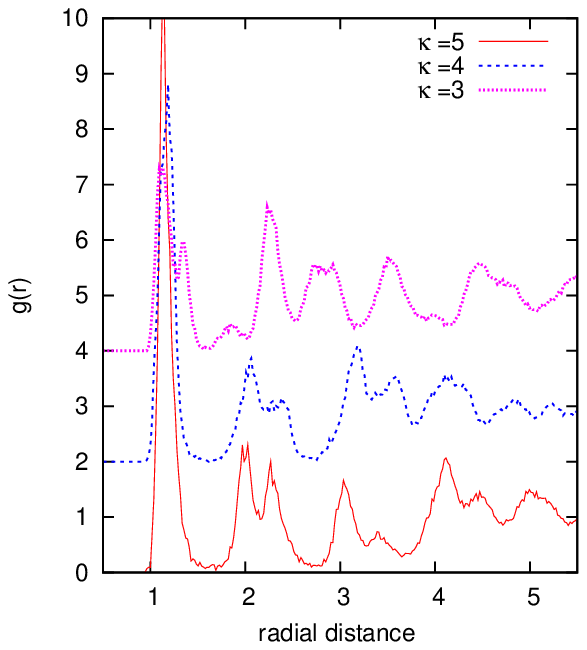}}
\subfigure[\label{fig:evo_tilt}]{\includegraphics[scale=0.9]{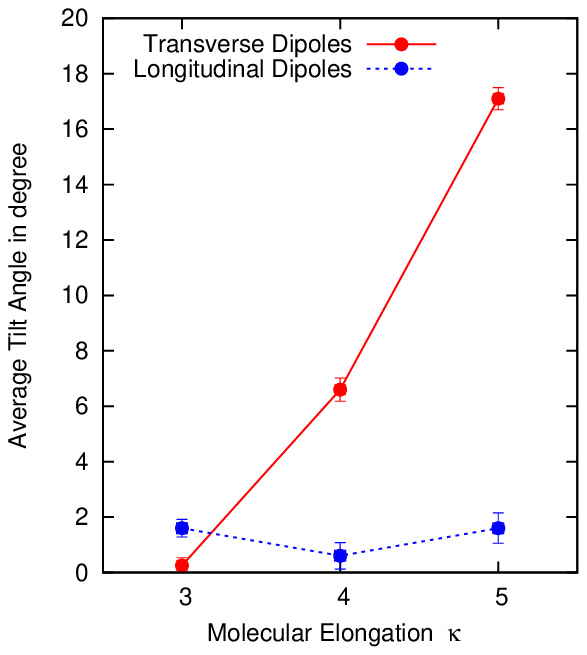}                 }
\caption{\label{fig:wide}(color online)(a) radial distribution functions for systems of molecules with two terminal longitudinal dipoles,(b) radial distribution functions for systems of molecules with two terminal transverse dipoles [ The zero of g(\(r^{}\)) on the vertical scales have been shifted for clarity ] (c) Schematic evolution of the average tilt angle  \(\langle \theta \rangle\) as a function
 of elongation \(\kappa\) for systems of molecules with two terminal dipoles .}
\end{figure*}

We have generated well equilibrated tilted smectic phases for three different elongations \(\kappa=3,4,5\) with two different dipolar orientations \(\phi=0^{\circ} \mbox{ and }90^{\circ}\) in order to explore separately the roles of a longitudinal component and a transverse component in producing tilted phases. From the snapshots and various distributions functions as described, we analyzed the phase structures. The structure of the phases were changed significantly over different elongations. 

In case of mesogens having longitudinal dipoles, the GB interaction plays dominant role in the isotropic phases and as we increase the pressure along the isotherms, the dipolar energy makes larger jump than the GB energy in reaching smectic phases. The energy distribution
in tilted phases show an interesting behavior over different elongations. For the shortest molecules( \( \kappa=3 \)),the GB energy
remains stronger than the dipolar energy and as we increase elongation they become comparable at \(\kappa=4\mbox{ and }5\). The amount of tilt
in shortest molecules \(\langle \theta \rangle=1.6^{\circ}\) decreases to \(\langle \theta \rangle=0.6^{\circ}\) in the \(\kappa=4\) system. The longest(\(\kappa=5\)) mesogens show \(\langle \theta \rangle=1.6^{\circ}\). The snapshots of the phases are shown in Fig.1 and the corresponding radial distribution functions are shown in Fig.\ref{fig:evo_tilt1}. We see that the amount of interdigitation in the smectic phases decreases with elongation showing an interdigitated phase at \( \kappa=3 \).

We now discuss the tilted structures obtained due to the effects of two transverse dipoles. In this case, the biaxiality comes
exclusively from the presence of the dipoles. Biaxial smectic phases are found
for all the three elongations. For \(\kappa=3\) we obtained an orthogonal biaxial phase and for other higher elongations 
we found tilted biaxial phases where the amount of tilt increases with \(\kappa\) as shown in the Fig.2. The contribution
 to the total energy is alaways dominated by the dipolar interaction in these phases and the dominance increases with the
 elongation. The average tilt order parameter as measured gives \(\langle \theta \rangle\approx 0^{\circ}\) for \(\kappa=3\) 
and more interestingly \(\langle \theta \rangle\approx 6.6^{\circ}\) and \(17^{\circ}\) for systems with \(\kappa=4\) and 5 respectively.
We may infer that longer dipolar separation gives larger torque arising out from larger dipole moments, which is responsible for giving significant change in the 
magnitude of tilt never found before. 
The values of the biaxial order parameter in the smectic phases are 
\(  \langle R_{2,2}^{2}\rangle\)=0.87,0.83 and 0.79 respectively for  \(\kappa=3,4 \mbox{ and }5\) systems.
The radial distribution function shows a crystalline nature in all three elongations at higher pressures as shown in Fig.\ref{fig:evo_tilt2}. In all the tilted phases, the smectic structure were never interdigitated
but show some difference in local hexagonal ordering as shown in Fig.2(d). Again the tilt angle increased from \(\langle \theta \rangle\approx\)  \(11^{\circ}\) to \(17^{\circ}\) in smectic phases 
as we increase the pressure \(P^{*}=0.45\) to \(P^{*}=1.5\) in systems of longest (\(\kappa=5\)) molecules. We noticed that the attempt to increase dipole moment by increasing \(\mu^{*}\)
 to a value greater than 1.1 fails due to enhanced probability of dimer formation.
We have observed the presence of Nematic Phase in these systems at different state points as a part of our ongoing work.\\

Our NPT simulation studies show that the increasing dipolar separation as a result of
molecular elongation can effectively give rise to large tilt in layered liquid crystalline phases
generated by GB molecules with two terminal transverse anti parallel dipole moments whereas it has less significant 
and reverse role in case of molecules with two longitudinal dipoles.
 Experminental evidence for tilted smectic phases were earlier reported for compounds 
with two or more lateral dipole moments \cite{jon1,jon2}. Our study is successful in gaining insights into the molecular origin 
of tilted phases by showing that terminal dipoles having longer separation length coupled with suitable orientation 
can bring large tilt to the liquid crystal phases.\\



T.K. Bose gratefully acknowledges the support of CSIR, India for providing Junior Research Fellowship via sanction no. 09/028(0794)/2010-EMR-I. The graphics software QMGA \cite{jon20} was used for producing snapshots of the phases. This work 
was supported by UGC-UPE scheme of the University of Calcutta.



\bibliography{PRL}

\end{document}